\documentclass[12pt,preprint]{aastex}
\newcommand{\HI}{H$^0$}
\newcommand{\HeI}{He$^0$}

\begin{document}
\title{Trajectories and Distribution of Interstellar Dust Grains in the
Heliosphere}

\author{Jonathan D. Slavin}
\affil{Harvard-Smithsonian Center for Astrophysics, MS 83, 60 Garden Street,
Cambridge, MA 02138}

\author{Priscilla C. Frisch}
\affil{University of Chicago, Department of Astronomy and Astrophysics, 5460
S.\ Ellis Avenue, Chicago, IL 60637}

\author{Hans-Reinhard M\"uller}
\affil{Department of Physics and Astronomy, Dartmouth College, Hanover, NH
03755}

\author{Jacob Heerikhuisen, Nikolai V. Pogorelov}
\affil{Department of Physics and Center for Space Physics and Aeronomic
Research, University of Alabama, Huntsville, AL 35899}

\author{William T. Reach}
\affil{Universities Space Research Association, MS 211-3, Moffett Field, CA
94035}

\and

\author{Gary Zank}
\affil{Department of Physics and Center for Space Plasma and Aeronomic
Research, University of Alabama, Huntsville, AL 35805}

\begin{abstract}
The solar wind carves a bubble in the surrounding interstellar medium (ISM),
known as the heliosphere.  Charged interstellar dust grains (ISDG)
encountering the heliosphere may be diverted around the heliopause or
penetrate it depending on their charge-to-mass ratio.   We present new
calculations of trajectories of ISDG in the heliosphere, and the dust
density distributions that result.  We include up-to-date grain charging
calculations using a realistic UV radiation field and full 3-D
magnetohydrodynamic fluid $+$ kinetic models for the heliosphere.  Models with
two different (constant) polarities for the solar wind magnetic field (SWMF)
are used, with the grain trajectory calculations done separately for each
polarity.  Small grains $a_\mathrm{gr} \lesssim 0.01$ \micron\ are completely
excluded from the inner heliosphere.  Large grains, $a_\mathrm{gr} \gtrsim
1.0$ \micron\ pass into the inner solar system and are concentrated near the
Sun by its gravity.  Trajectories of intermediate size grains depend strongly
on the SWMF polarity.  When the field has magnetic north pointing to ecliptic
north, the field de-focuses the grains resulting in low densities in the inner
heliosphere, while for the opposite polarity the dust is focused near the Sun.
The ISDG density outside the heliosphere inferred from applying the model
results to \emph{in situ} dust measurements is inconsistent with local ISM
depletion data for both SWMF polarities, but is bracketed by them. This result
points to the need to include the time variation in the SWMF polarity during
grain propagation.  Our results provide valuable insights for 
interpretation of the \emph{in situ} dust observations from \emph{Ulysses}.
\end{abstract}


\section{Introduction \label{sec:intro}}
The Sun is traveling through a magnetized, low density, partially ionized
interstellar cloud \citep{Frisch_etal_2011}. The relative motion of the Sun
and the cloud and the outflowing solar wind combine to produce the bow shaped
heliosphere.  The heliosphere consists of three distinct regions: the inner
heliosphere in which the solar wind plasma expands almost freely, the inner
heliosheath, which contains shocked solar wind plasma, and the outer
heliosheath, which has heated and decelerated interstellar medium (ISM)
plasma.  If the interstellar magnetic field were small enough so that the
magnetosonic speed in the surrounding ISM was below the relative speed of the
solar system and the ISM, there would be a second shock marking the transition
from the undisturbed ISM to the outer heliosheath.  Data from a variety of
sources including \emph{Voyager 1} and \emph{Voyager 2} and the
\emph{Interstellar Boundary Explorer} (IBEX) point to a relatively large
magnetic field, $B \approx 3\,\mu$G oriented at a fairly large angle ($\sim
48\degr$) relative to the upwind direction \citep[and references
therein]{McComas_etal_2012}.  Therefore, the surrounding ISM, known as the
Local Interstellar Cloud (LIC) is now thought to move subsonically relative to
the Solar System and so there is no outer bow shock but rather a continuous
transition beyond the heliopause that nevertheless results in heating of the
plasma and an increase in the density and magnetic field strength of the ISM.
The heliopause, a contact discontinuity, is the boundary between inner and
outer heliosheath, and the solar wind termination shock marks the transition
from the inner heliosphere to the inner heliosheath\footnote{See
\citet{Zank_1999} for a review of the interaction between the solar wind and
interstellar material for cases where the heliosphere has only one shock,
versus both a termination shock and a bow shock.}.

Though the above picture would seem to imply a clean separation of the solar
wind plasma and interstellar medium, in reality there are a variety of
interactions that couple components of the ISM and the solar wind throughout
the heliosphere.  The existence of pickup ions (PUI) is one manifestation of
this coupling, since their primary source is the population of neutral
interstellar atoms that are ionized by charge exchange or photoionization in
the inner heliosphere and subsequently picked up by the solar wind
\citep[e.g.][]{Gloeckler+Fisk_2007,Zank_1999}.  Interstellar dust in the solar
system, observed by the \emph{Ulysses}, \emph{Galileo} and \emph{Cassini}
spacecrafts, is another example of an interstellar population modified by
interaction with the heliosphere
\citep[e.g.][]{Morfill+Gruen_1979,Frisch_etal_1999,
Landgraf_etal_2000,Altobelli_etal_2007,Krueger_etal_2010}\footnote{See
\citet{Mann_2010} for a review of the interactions between interstellar dust
grains and the heliosphere.}.  The interaction between the interstellar dust
and the heliosphere, which depends on the solar wind magnetic field and grain
charging by the surrounding plasma and UV radiation field, is in many ways
more complex than the PUI interactions.  The purpose of this study is to model
the distributions of interstellar dust grains propagating into and through the
heliosphere from the undisturbed ISM.

Interstellar dust (ISD) grains outside of the heliosphere span a broad range
of sizes from Angstrom sizes (for polycyclic aromatic hydrocarbon dust) to
several microns for the largest grains in cold molecular cloud cores
\citep{Pagani_etal_2010}.  The charge-to-mass ratios calculated for the larger
ISD grains imply gyroradii that are large compared to the size of the
heliosphere. The smaller grains, however, are tightly coupled to the
magnetic field, since their gyroradii are fractions of an AU, and are swept
around the heliosphere along with the magnetic field.  For medium sized grains
with gyroradii on the order of a few AU or so it is unclear what to expect.
For these grains one needs detailed calculations of their trajectories to
assess their level of density depletion or enhancement in the heliosphere and
its spatial dependence.

Among the mysteries regarding the ISD observed inside of the solar system is
to what extent the observed grain size distribution is representative of the
undisturbed size distribution in the surrounding interstellar cloud. This
point is of particular interest because the size distribution is so unusual
compared to that inferred for the ISM in general from extinction of starlight,
infrared emission and other evidence \citep{Draine_2009}. The observed grains
span a range from $\sim 0.05 - 1$ \micron\ in size
\citep{Grun_etal_1994,Grun+Svestka_1996} as compared to the canonical model
of \citet[][MRN]{Mathis_etal_1977} for the interstellar dust size distribution
which goes from 0.01 to 0.25 \micron\ (for silicates, with the lower end quite
uncertain).  Interstellar dust models that simultaneously model grain infrared
emissivity and extinction data, while satisfying the abundance constraints,
require grain sizes of up to 0.8 \micron, using a range of possible grain
compositions and a mix of carbonaceous and silicate grains
\citep{Zubko_etal_2004}.  The amount of mass in the observed dust, if typical
depletion of elements into grains is assumed, is marginally inconsistent with
cosmic abundances (depending on the abundance set assumed) given the inferred
interstellar gas density for the cloud \citep{Slavin+Frisch_2008}.  As we
discuss below, however, the observed dust mass and size distribution is
unlikely to closely match that in the undisturbed ISM outside of the
heliosphere and so determining how to infer that initial size distribution is
an important motivation for this work.

Trajectories of interstellar grains through the inner heliosphere have been
calculated previously, and it was shown that the interaction of the charged
grains with the solar wind magnetic field produces trajectories that depend on
the phase of the solar magnetic activity cycle 
\citep[e.g.][]{Landgraf_2000,Grogan_etal_1996,Landgraf_etal_2003,
Sterken_etal_2012}.  Common properties of these models are the analytical
treatment of the solar cycle variations of the solar wind magnetic field and a
computational domain that includes a uniform source region 40--50 AU from the
Sun.  All these models show that grains
alternately focus or de-focus with respect to the ecliptic plane, depending on
the solar cycle phase.  None of the previous models include the asymmetries of
the full 3-D global heliosphere revealed by the \emph{Voyager} termination
shock crossings \citep{Stone_etal_2008}, the IBEX data including the ribbon of
energetic neutral atoms \citep{McComas_etal_2009}, and the offset between the
inflow direction of interstellar \HeI\ and \HI\ into the heliosphere
\citep{Lallement_etal_2010}.  Most calculations assume a uniform grain source
upwind of the Sun and without regard for the structure outer heliosphere.
\citet{Czechowski+Mann_2003} have investigated the dynamics of $\sim 0.01$
\micron\ grains in the heliosheath regions and found that such grains may
stream along the heliosphere flanks, and in addition trace variations in the
plasma density.  Recently, \citet{Sterken_etal_2012} have presented a
parameter study of grain trajectories in the time-variable inner heliosphere,
within $\sim 30$ AU of the Sun for a uniform density grain source.  

Our simulations of interstellar grains interacting with the heliosphere make
use of global models that represent the heliosphere during either the focusing
or de-focusing phase of the solar magnetic activity cycle, and provide the
first study of the distribution of interstellar dust grains in the
magnetically distorted global heliosphere.  The heliosphere model that we use
is a fully 3-D MHD-kinetic model (MHD for the ions and kinetic for the
neutrals) described by \citet{Pogorelov_etal_2008}.  This model was created
with the aim of matching all the available data on the heliosphere including
the asymmetry indicated by the crossings of the termination shock by
\emph{Voyager} 1 and 2 \citep{Stone_etal_2008}, the H deflection plane data
\citep{Lallement_etal_2005,Lallement_etal_2010}, and the Ulysses measurement
of the relative Sun-LIC velocity \citep{Witte_2004}.  It has been highly
successful in predicting the direction of the interstellar magnetic field
(ISMF) as inferred from data from \emph{IBEX}. The region of the sky with
enhanced energetic neutral atom flux detected by \emph{IBEX} is known as the
``ribbon'' \citep[discovered by][]{McComas_etal_2009} and models for its
production predict that its location is sensitive to the orientation of the
interstellar magnetic field.  The center of the ribbon arc, at $\ell=33^\circ
\pm 4^\circ$, $b = 55^\circ \pm 4^\circ$, is located $\sim 17\degr$ from the
direction of the ISMF used in our models, and $48^\circ$ from the interstellar
gas flow direction.\footnote{Although \emph{a posteriori} the $17^\circ$
offset appears to be a significant disagreement, it is remarkable that the
heliosphere models created prior to the \emph{IBEX} observations are in such
good agreement with the ribbon location, if one assumes that sightlines
perpendicular to the ISMF draping over the heliosphere form the locus of
points on the sky containing the unpredicted ribbon
\citep{Schwadron_etal_2009}.} The angle between the ISMF and the gas flow
produces a magnetically distorted heliosphere that is incorporated into the
models we use. The bending of the interstellar magnetic field lines as they
are wrapped around the asymmetric heliosphere has important consequences for
the trajectories, particularly for grains with sizes near 0.1 \micron.

While our simulations provide the first study of the interactions between
interstellar dust and the global 3-D heliosphere, one of our conclusions is
that the trajectories of sub-micron grains that penetrate the inner
heliosheath are sensitive to the solar wind magnetic field polarity. As a
result an accurate model of the propagation of these grains through the inner
heliosheath regions will require a time-variable heliosphere that takes into
account the variation of the SWMF because of the 22 year solar magnetic
activity cycle.

\section{Methods} \label{sec:method}

\subsection{Heliosphere Model \label{sec:heliomodel}}

A big step toward constraining MHD models of the heliosphere was taken when
the \emph{Voyager 1} and \emph{Voyager 2} spacecraft encountered the
termination shock at different distances. These data, together with the $\sim
7.8\degr$ offset between the upwind interstellar \HI\ and \HeI\ directions
\citep[the so-called hydrogen deflection plane,
HDP][]{Lallement_etal_2005,Lallement_etal_2010}, and the $48\degr$ angle
between \HeI\ upwind direction and ISMF given by the center of the \emph{IBEX}
ribbon, have established the asymmetry of the global heliosphere.  The
magnetically distorted heliosphere is confirmed by \emph{IBEX} ENA maps
showing that the tail of the heliosphere is offset from the downwind gas
direction by $\sim 40\degr$ \citep{Schwadron_etal_2011}, and by ENA fluxes
from the polar regions that reveal that the inner heliosheath in the north is
thicker by $\sim 27$ AU than in the south \citep{Reisenfeld_etal_2012}.

\begin{figure}[!ht]
\plotone{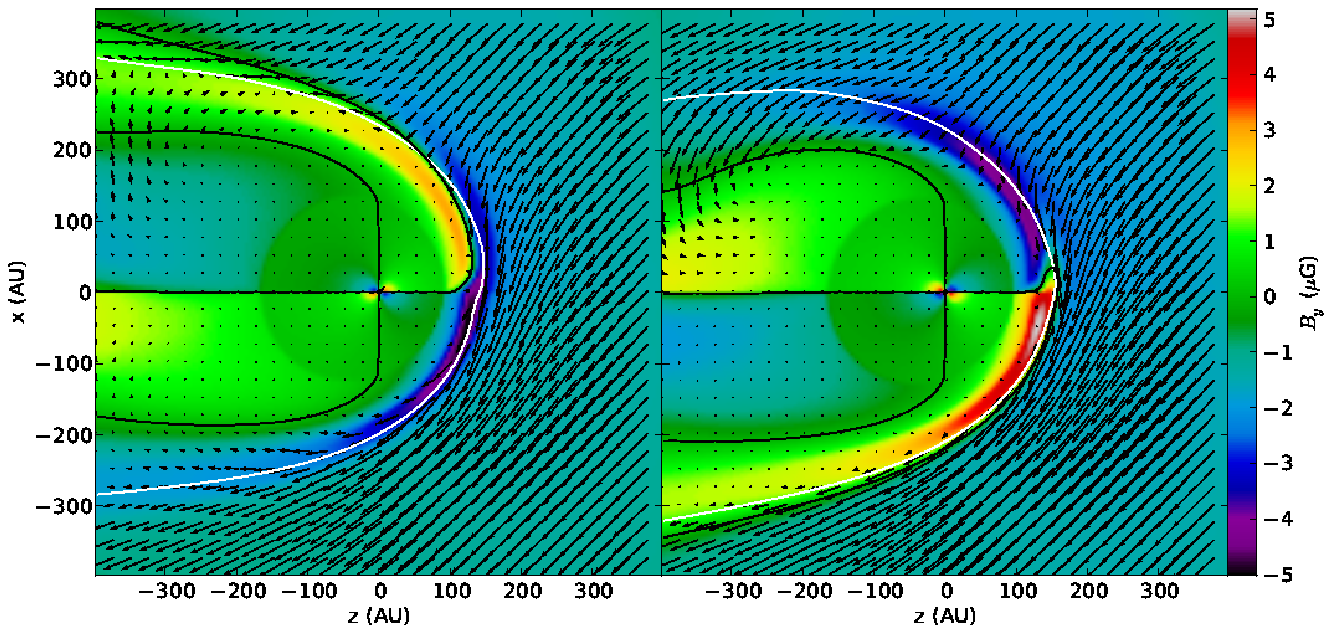}
\caption{Magnetic field for the heliosphere models used with different solar
wind magnetic field polarities, magnetic north positive or de-focusing (left)
and magnetic north negative or focusing (right).  The colors indicate the
strength of the $y$ component of the $B$ field (out of the page) while the
arrows indicate the strength and direction of the field projected onto the
$x-z$ plane.  The white curves indicate the shape of the heliopause, while the
black lines separate regions of opposite polarity of $B_y$. The different
polarities of the field lead to different shapes of the heliosphere because of
magnetic reconnection near the heliopause.
\label{fig:comp_Bfields}}
\end{figure}

The heliosphere models that we use in our dust trajectory calculations,
described more fully by \citet{Pogorelov_etal_2008} and
\citet{Heerikhuisen_etal_2006, Heerikhuisen_etal_2008b}, are mixed
magnetohydrodynamic/kinetic models with the ions treated as a fluid while
neutrals are treated kinetically using a statistical approach, with both
particle species coupled self-consistently by charge exchange.  These models
have been successful in explaining the location of the IBEX Ribbon of
energetic neutral atoms \citep[which is centered $\sim 17\degr$ away from
the interstellar field direction in our models ][]{McComas_etal_2009} and the
north-south asymmetry of the heliosphere, and are consistent with the offset
between the inflows of \HI\ and \HeI\ into the heliosphere.  The coordinate
system used in the calculations is centered on the Sun and has the $z-$axis
pointed toward the ecliptic longitude of the upwind direction, i.e.\ ecliptic
longitude $\lambda = 259\degr$, but in the ecliptic plane, ecliptic latitude
$\beta = 0$, which makes it point about $5\degr$ away from the upwind
direction. The $x-$axis is perpendicular to the ecliptic plane and the
$y-$axis, completing the right-handed coordinate system, lies in the ecliptic
plane.  The particular model we use in this work has an initial field
direction in the undisturbed ISM of $(\lambda,\beta) = (61\degr, -30\degr)$
(ecliptic coordinates, see Table 1 for the direction in galactic coordinates).
The projection of the B-field onto the $x-z$ plane can be seen in Figure
\ref{fig:comp_Bfields} where we compare the two heliosphere models that we use
with different solar wind magnetic field polarities.  As the interstellar
plasma encounters the heliosphere, the field is draped around it creating a
curvature that tends to divert the small grains around the heliosphere.  As
discussed below, the grains tend to be positively charged by the solar UV
radiation once they get within $\sim 100$ AU of the Sun.

\begin{deluxetable}{lcc}
\tablecolumns{3}
\tablewidth{0pc}
\tablecaption{Boundary Conditions Used in the Heliosphere Models\label{tab:BCs}}
\tablehead{\colhead{Quantity} & \colhead{Solar wind\tablenotemark{a}} & 
\colhead{ISM\tablenotemark{b}}}
\startdata
$n_p$ (cm$^{-3}$) & 7.4 & 0.06 \\
$T$ (K) & $10^5$ & 6,527 \\
$B$ ($\mu$G) & 37.5\tablenotemark{c} & 3.0, 3.6\tablenotemark{d} \\
$B$ direction\tablenotemark{e} & \nodata & $\ell, b = 203\degr,-38\degr$  \\
$v$ (km s$^{-1}$) & 450 & 26.4 \\
$v$ upwind & \nodata & $\ell, b = 4\degr,15\degr$ \\
\enddata
\tablenotetext{a}{Boundary conditions at 1 AU.  These conditions are
then advected to innermost edge of the 3-D grid, $R = 10$ AU, solving for the
velocity, density and temperature changes along the way.}
\tablenotetext{b}{Values far upstream of the heliosphere ($\sim 1200$ AU).}
\tablenotetext{c}{Total B field (mostly toroidal or azimuthal) near the
ecliptic. Note that the field passes through a null in the ecliptic plane
because of the change in polarity between the northern and southern
hemisphere.}
\tablenotetext{d}{For the de-focusing and focusing solar wind magnetic
field cases respectively.}
\tablenotetext{e}{The solar wind B field has the standard Parker spiral
morphology. Note that this ISMF direction is $\sim 17\degr$ from the center of
the IBEX ribbon at $\ell,b = 33\degr,55\degr$, which is thought to be an
indicator of the direction of the ISMF.}
\end{deluxetable}

The solar wind in the heliosphere models has a radial velocity at 1 AU of 450
km s$^{-1}$ and density of 7.4 cm$^{-3}$ and has a Parker spiral magnetic
field with strength 37.5 $\mu$G.  These boundary conditions are then advected
to the inner edge of the 3-D grid at $R = 10$ AU.  This results in values in
the innermost parcel of grid of $v_r \approx 454$ km s$^{-1}$, $n_p \approx
0.07$ cm$^{-3}$ and $B_\phi \approx 3.3\,\mu$G (near the ecliptic). The
temperature evolution in the wind is strongly affected by the interactions
between interstellar neutrals and the solar wind ions that produce the pickup
ions, and is anisotropic at 10 AU with values ranging from 2800 K to 14,000 K
with a mean of 3800 K.  These and other boundary conditions in the heliosphere
models are summarized in Table \ref{tab:BCs}.  It has long been known that the
large scale solar magnetic field changes polarity every solar cycle \citep[and
references therein]{Howard_1977} and these polarity changes propagate outward
with the solar wind.  We explore models with both different polarities for the
solar wind magnetic field in this work, the focusing polarity that has north
pole negative (i.e.\ north ecliptic pole is south magnetic pole, a.k.a.\ the
$A^-$ configuration) and de-focusing that has north pole positive ($A^{+}$
configuration). The reason that the field with north pole negative polarity is
focusing for the dust is that, as we discuss further below, the grains are
positively charged.  Thus $q \vec v \times \vec B$ points down toward the
ecliptic for grains in the north (i.e. above the ecliptic plane) and points up
toward the ecliptic for grains in the south. Note that the models we use are
steady flow models and so the solar wind magnetic polarity does not evolve.
The interstellar medium before the encounter with the heliosphere is assumed
to be flowing with speed 26.4 km s$^{-1}$ relative to the Sun and have a total
proton density of 0.06 cm$^{-3}$ and temperature of 6527 K.  The neutral H
density in the ISM is assumed to be 0.15 cm$^{-3}$.

The heliosphere models use a 3-D grid in spherical coordinates (with the polar
axis directed towards the longitude of the heliosphere nose), with an inner
boundary (radius) at 10 AU and an outer boundary at 1200 AU with 224 zones in
the $r$ direction (logarithmically spaced), 144 zones in $\theta$ and 83 in
$\phi$.  The heliosphere model results include the electron temperature,
proton density, magnetic field and velocity over this grid. The grain density
grid used is a 3-D Cartesian grid with a spacing of 5 AU in each direction and
extending from $-400$ AU to $+400$ AU in each dimension. Note that the
computation of the grain trajectories starts farther upwind (900 AU) but the
densities are only recorded once the grains enter the density grid.
Inside of 10 AU, the dynamical variables are assumed to have their values at
the inner edge of the grid.

\subsection{Dust Model \label{sec:dustmodel}}

We focus on compact olivine-type silicate grains in our models in this paper.
\citet{Slavin+Frisch_2008} have shown that absorption line data toward
$\epsilon$ CMa provide strong evidence that C is not depleted from the gas
phase in the LIC, which appears to be the source of the gas flowing into the
solar system \citep{Frisch_etal_2011}.  The depletion pattern of refractory
elements in the LIC suggest a grain composition similar to that of olivines,
which appear to be widespread in the local ISM \citep{Frisch_etal_2011}.
\citet{Kimura_etal_2003a} conclude that C in the local ISM is slightly
depleted, but exclude the low \ion{H}{1} column density lines of sight in this
determination.  Thus their method includes mostly gas that is not in the LIC
but rather in other clouds.  
The differences between silicate and carbonaceous grains for the purposes of
this paper include the size-to-mass ratio (i.e.\ the density of the grain
material), the charging properties and the grain optical properties (which
help determine the value of $\beta$, see below). If future work indicates C
depletion in the LIC or the presence of carbonaceous interstellar grains in
the heliosphere, a study of such grains would be warranted, though we do not
include them in the present study.
The solid material density of the grains that we use
is 3.3 g cm$^{-3}$, as appropriate for olivine silicates, in the results
presented in this paper. It has been suggested \citep[e.g.,][]{Mathis_1996} that
some or even most interstellar grains may have a porous, or fluffy structure
leading to a considerably lower material density.  The first fluffy grain
models have been ruled out, however based on the excessive amount of infrared
emission they would produce \citep{Dwek_1997}. It is currently unclear what
fraction of interstellar grains might be fluffy.  We intend to explore
alternative grain types in future work.

\begin{figure}[ht!]
\plotone{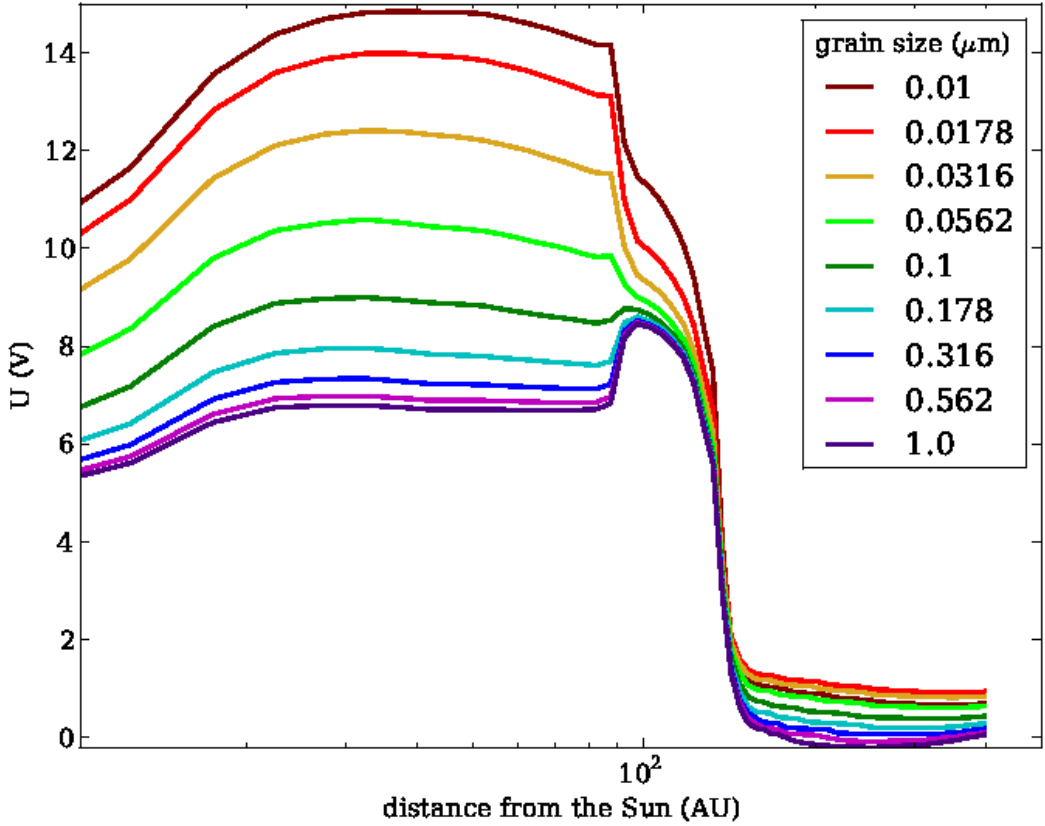}
\caption{Grain potential (in volts) vs.\ distance from the Sun (in the
upstream direction) for a range of grain sizes.  This illustrative calculation
was done for grains lying along the $z-$axis (5\degr\ from the upstream
direction, see text for details about the coordinate system used).  The
potential varies as a function of the strength of the radiation field (with
the solar field going as $1/r^2$), but is even more strongly affected in these
curves by the plasma temperature.  The rise in the potential going outward is
thus related to the rise in temperature, which in turn is caused by heating of
the solar wind by the pickup ions. The temperature is high in the heliosheath
(between 90 and 130 AU), strongly affecting the grain charge.  The effects of
relative gas-grain velocity are not included in this plot (but are in the
trajectory calculations) and tend to increase the potential, especially in the
inner heliosphere, by roughly 10\%.  Some previous modelers of grain
trajectories in the heliosphere have used a constant value of 5 V, which is
clearly a poor fit to our results with the exception of the largest grains in
the inner heliosphere.
\label{fig:U_vs_r}}
\end{figure}

\begin{figure}[ht!]
\plotone{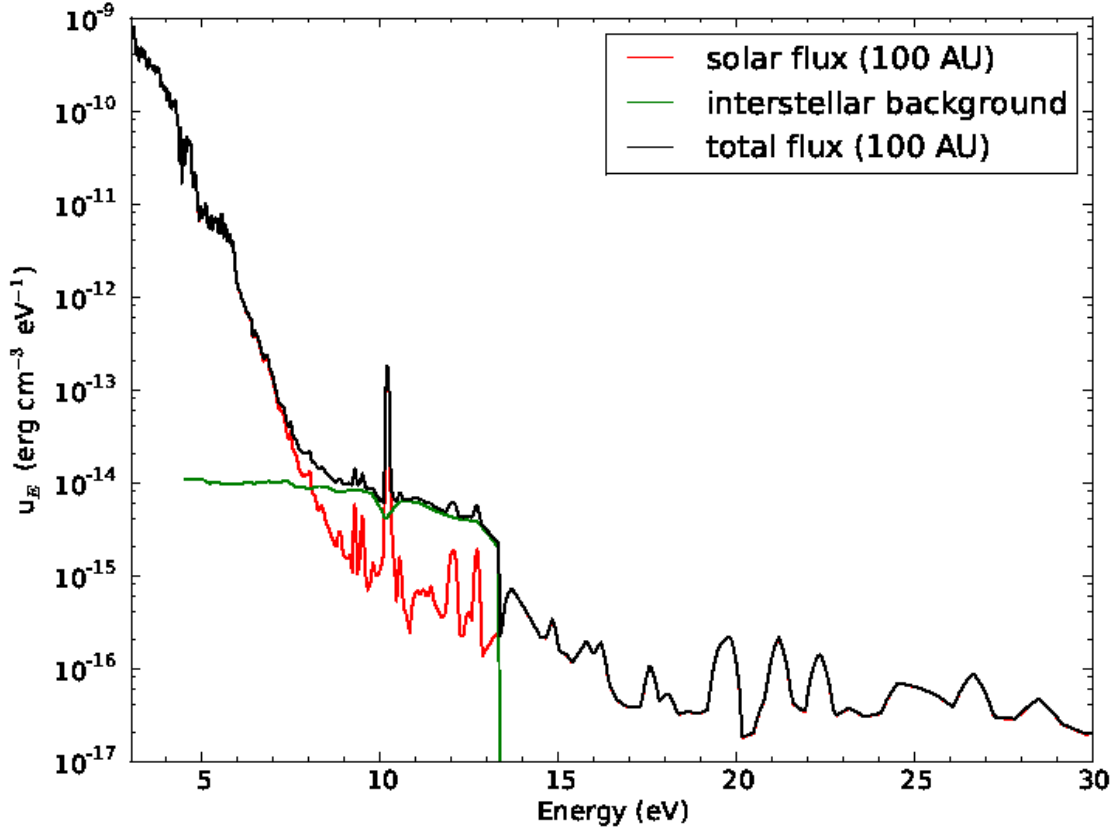}
\caption{Visible/UV/EUV spectral energy density at 100 AU distance
from the Sun.  The solar part of the spectrum is based on data from the
TIMED-SEE and SORCE missions, while the interstellar background portion is
based on the results of \citet{Gondhalekar_etal_1980}.  For charging of
silicate grains only the FUV/EUV part of the spectrum above 8 eV (the work
function for the grains) is important.  The visible and NUV ($\sim 3 - 6$ eV)
parts of the spectrum dominate for heating the grains because of the large
flux at those energies.  It is clear that the FUV interstellar background
radiation field is important at 100 AU, though it becomes less important as
the grains penetrate closer to the Sun.
\label{fig:solar_bg_spect}}
\end{figure}

The charging properties of interstellar grains are still quite uncertain.  In
our calculations we use the optical constants and photoelectric yields of
\citet{Weingartner_etal_2006}, making use of their code (kindly provided to us
by J.\ Weingartner), though we have made small modifications to include the
effect of gas-grain motion on the charging \citep[using methods described
in][]{Guillet_etal_2007}.  The main sources of charging for interstellar
grains in the heliosphere are electron impacts and sticking and photoelectric
charging by the FUV radiation field. In regions of hot gas ($T \gtrsim 10^5$
K), electrons have enough energy to cause the ejection of secondary electrons,
which tends to positively charge the grains.  The resultant grain potentials
that we find for the upstream direction towards the heliosphere nose are
illustrated in Figure \ref{fig:U_vs_r}.  The enhanced grain charging in the
inner heliosheath, originally discussed by \citet{Kimura+Mann_1999}, is
clearly visible.  We find that the grain potential is strongly position and
grain size dependent.  While the FUV field is dominated by solar radiation in
the inner heliosphere, the background interstellar radiation field dominates
in the farther regions of the outer heliosphere (see Figure
\ref{fig:solar_bg_spect}).  We use the UV radiation field of
\citet{Gondhalekar_etal_1980} for the background interstellar field
\citep[e.g. see the radiation longward of 912 \AA\ in Fig. 1
of][]{Slavin+Frisch_2008}.  This field corresponds to a UV flux, averaged over
grain photoelectric yield, of 0.74 times that of \citet{Draine+Salpeter_1979}.
For the solar field we use data from the Solar EUV Experiment (SEE) on the
NASA TIMED (Thermosphere Ionosphere Mesosphere Energetics and Dynamics,
\url[http://lasp.colorado.edu/see/l3\_data\_page.html]
{http://lasp.colorado.edu/see/l3\_data\_page.html}) mission which provides
daily EUV/FUV spectra based on solar observations and models and from the
SORCE (Solar Radiation \& Climate Experiment,
\url[http://lasp.colorado.edu/sorce/index.htm]
{http://lasp.colorado.edu/sorce/index.htm}) mission that provides better
visible/near UV data.

For the inclusion of the effects of radiation pressure and gravity we
calculate $\beta$, the ratio of the radiation pressure force to the
gravitational force as a function of grain size, 
\begin{equation}
\beta = -\frac{F_r}{F_g} = \frac{R^2}{G M_\odot m_\mathrm{gr}}
\frac{1}{c} \int F_\lambda \sigma_\mathrm{gr} Q_\mathrm{p}\,d\lambda,
\end{equation}
where $F_r$ is the radiation pressure force, $F_g$ is the gravitational force,
$R$ is the distance from the Sun, $m_\mathrm{gr}$ is the grain mass,
$F_\lambda$ is the flux (erg cm$^{-2}$ s$^{-1}$ \AA$^{-1}$) at the distance
$R$, $\sigma_\mathrm{gr}$ is the physical cross section of the grain and
$Q_\mathrm{p}$ is the radiation pressure efficiency factor.  
(For more details see discussions in \citet{Gustafson_1994} and
\citet{Sterken_etal_2012}.) Note that $R$ drops out since the flux goes as
$1/R^2$.  This calculation depends on the grain optical properties, through
$Q_\mathrm{p}$, which in turn depends on the absorption efficiency,
$Q_\mathrm{a}$, and scattering efficiency, $Q_\mathrm{s}$,
\begin{equation}
Q_\mathrm{p} = Q_\mathrm{a} + (1 - g) Q_\mathrm{s}
\end{equation}
where $g \equiv \langle\cos(\theta)\rangle$ is the scattering asymmetry
factor.  We calculate the radiation pressure efficiency using results and code
from \citet{Draine_2003}.  Using the solar spectrum and our assumed grain
density, $\rho_\mathrm{gr} = 3.3$ g cm$^{-3}$, we find that $\beta$ is less
than one for all grain sizes.  We note that $\beta$ is sensitive to the grain
density and composition \citep[Fig. 2 in ][]{Kimura_etal_2003b}, and $\beta$
exceeds 1 (net repulsion of grains) for low density (fluffy) grains.  There is
some observational evidence for repulsion of grains via radiation pressure,
primarily for small grains within 4 AU of the Sun \citep[Fig. 2 in
][]{Mann_2010}.  \citet{Sterken_etal_2012} model variations in $\beta$ and
show that $\beta$ is important only within several AU of the Sun (except for
downstream where the effects extend farther), in accord
with the observations.  Given our assumptions, however, the net effect from
radiation pressure and gravity on the grains is gravitational focusing though
the effect is minor throughout most of the heliosphere.  The peak value for
$\beta$ is 0.89 at $a_\mathrm{grain} \approx 0.2$ \micron\ and so gravity and
radiation pressure nearly balance and grains near that size experience little
effect from the Sun's gravity or radiation pressure. As we discuss below and
in \citet{Slavin_etal_2010}, in our models smaller grains are mostly excluded
from the inner heliosphere by the solar wind magnetic field, so that
uncertainties in $\beta$ do not strongly affect those results.  Since the
value of $\beta$ decreases as $1/a_\mathrm{grain}$ above 0.2 \micron, the
gravitational effects on the grain trajectories are greatest for the largest
grains.  These grains also have the smallest charge-to-mass ratios, so
gravitational effects play a dominant role in determining their trajectories
near the Sun.

\section{Grain Trajectory Calculations\label{sec:trajectory}}
We calculate the trajectory of grains by integrating the equation of motion,
\begin{equation}
\ddot{\vec{x}}_\mathrm{g} = -(1 - \beta) \frac{GM_{\sun}}
{|\vec{x}_\mathrm{g}|^3} \vec{x}_\mathrm{g} + \frac{q}{m}
((\dot{\vec{x}}_\mathrm{g} -  \vec{v}_\mathrm{gas}) \times \vec{B}),
\end{equation}
where $\vec{x}_\mathrm{g}$ is the grain position relative to the Sun, $q$ is
the grain charge, $m$ is the grain mass and $\vec{B}$ is the magnetic field.
We also include both the direct drag from gas-grain collisions and plasma drag
in our calculations but have found these effects negligible on the scale of
the heliosphere.  At each point in the integration, the charge is found under
the assumption of equilibrium charging based on the radiation field, plasma
density, temperature and gas-grain relative velocity at the position of the
grain.  The equilibrium assumption has been tested and found to be very good.
We ignore the quantization of charge in our calculations, which, for small
grains, introduces a distribution of possible charge states for any given set
of charging conditions \citep[see, e.g.][]{Weingartner+Draine_2001}.  In
reality the charge distribution will be sampled rapidly by a grain because the
charging rates all have short time scales, $\tau \lesssim
10\,\mathrm{s}/a(\mu\mathrm{m})^2$.  As a result the grain charge will
vary with time but the trajectory will tend to hover around that for the
equilibrium assumption.  Over many grains we expect the net effect of this
time varying charge to be small (especially considering the uncertainties in
grain charging) and we ignore it in our calculations presented here.

The other quantities that affect the trajectory of a grain, namely the
electron and ion density, magnetic field strength and direction, and gas
velocity, are taken from the pre-calculated heliosphere models discussed above
(see Table \ref{tab:BCs}).  The quantities are found via 3-D interpolation of
the grid-based model results.  We start the grains at a distance of 900 AU
from the heliosphere, moving mainly with the gas and with charge appropriate
to the conditions.  Because grains may be substantially accelerated in the
ISM, simply from turbulence present within the cloud, we introduce an initial
gyration of the grains around the magnetic field.  This initial relative
velocity is assumed to be 3 km s$^{-1}$ and is perpendicular to the local
magnetic field (i.e.\ pitch angle of $90\degr$) but otherwise randomly
directed within that plane.  This choice was made to be roughly consistent
with recent work on grain acceleration by turbulence by \citet{Yan_etal_2004}
in which they find that the grains are accelerated to the gas turbulent
velocity or more and have large pitch angles.  The turbulent velocity of the
gas in the LIC is generally found to be $\lesssim 3$ km s$^{-1}$
\citep{Redfield+Linsky_2008}.  The effect of adding this initial gas-grain
motion is not large and mainly serves to smooth out the signature of the
initial grid of grain starting points in our final results for grain density
presented below.

For each grain size we calculate trajectories of over 10$^6$ grains in an
initial $1001\times1001$ grid spaced 1 AU apart. We use the freely available
VODE code\footnote{available at
\url[https://computation.llnl.gov/casc/odepack/odepack_home.html]
{https://computation.llnl.gov/casc/odepack/odepack\_home.html}} to carry out
the numerical integrations. To calculate the grain space density we set up a
grid that is $160\times160\times160$ with each cubic voxel 5 AU on a side.
After each (fixed) output timestep the count of grains within the voxel in
which the grain is located is incremented. With our chosen output timestep we
get $\sim 60$ points per voxel (per grain size) in the region where the grain
density has yet to be significantly changed from its initial density.  To
normalize the density to its physical value we need to multiply by the actual
assumed density of grains in the ambient ISM.  This in turn depends on the
gas-to-dust mass ratio, the grain material density (given above) and the
initial grain size distribution.  Ideally we could go from the observed
properties of ISD in the heliosphere to infer the properties of the dust in
the ambient ISM, however for completely excluded grains this is obviously not
possible and for regions of low density, we are limited by our statistics.

\section{Results \label{sec:results}}
\subsection{Dust Density Distributions \label{sec:density}}
\begin{figure}[ht!]
\plotone{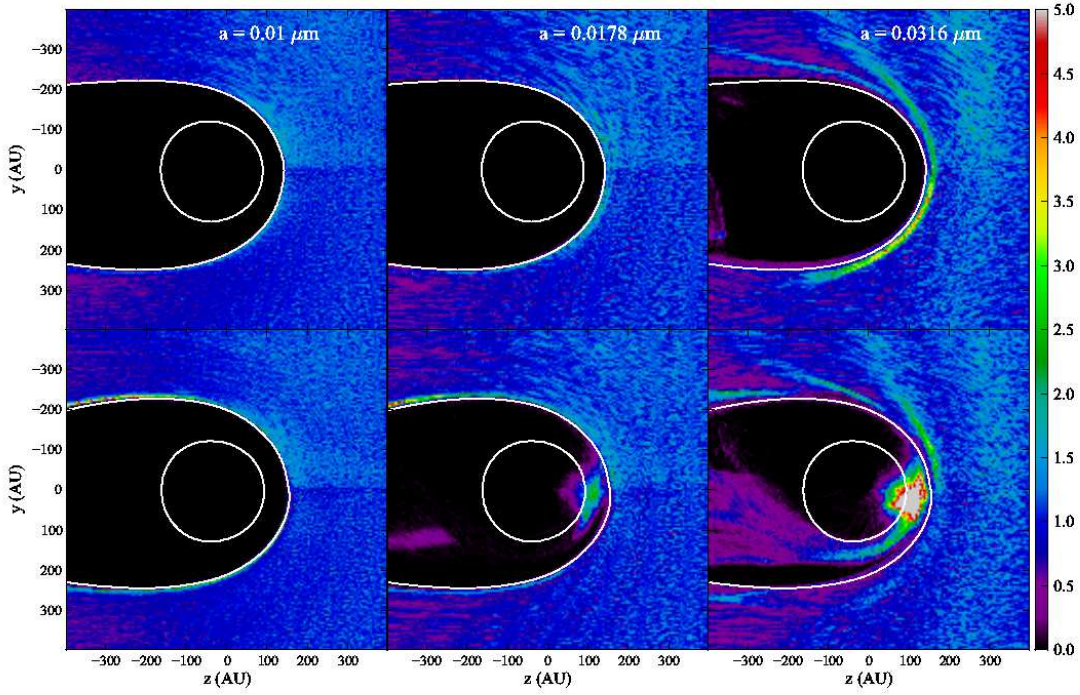}
\caption{Dust density distributions in slices parallel to the ecliptic plane
for 0.01, 0.0178 and 0.0316 \micron\ grains (as labeled).  The top row is for
the de-focusing SWMF polarity and the bottom row is for the focusing polarity.
The color scale indicates the density relative to the ambient interstellar
dust density for that grain size.  The white curves indicate the termination
shock (inner, nearly circular curve) and the heliopause (outer curve).  The
striping in the image is an artifact of the initial grid of trajectory
starting positions.}
\label{fig:ddens_group1_x}
\end{figure}

\begin{figure}[ht!]
\plotone{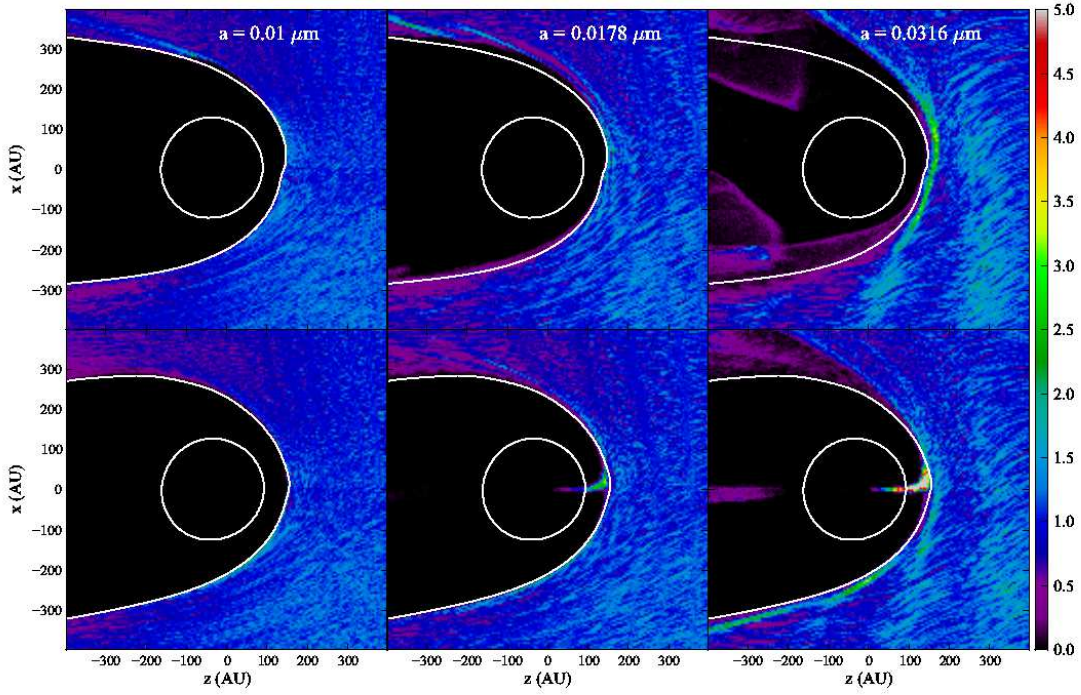}
\caption{Same as Figure \ref{fig:ddens_group1_x} except for slices
perpendicular to the ecliptic plane and in the plane of the inflow.  Upstream
is to the right.  As in Figure \ref{fig:ddens_group1_x} the top row is for the
de-focusing SWMF and the bottom row is for the focusing polarity. Also as in
Fig. \ref{fig:ddens_group1_x} the inner white curve indicates the location of
the termination shock while the outer white curve shows the heliopause
location.}
\label{fig:ddens_group1_y}
\end{figure}

\begin{figure}[ht!]
\plotone{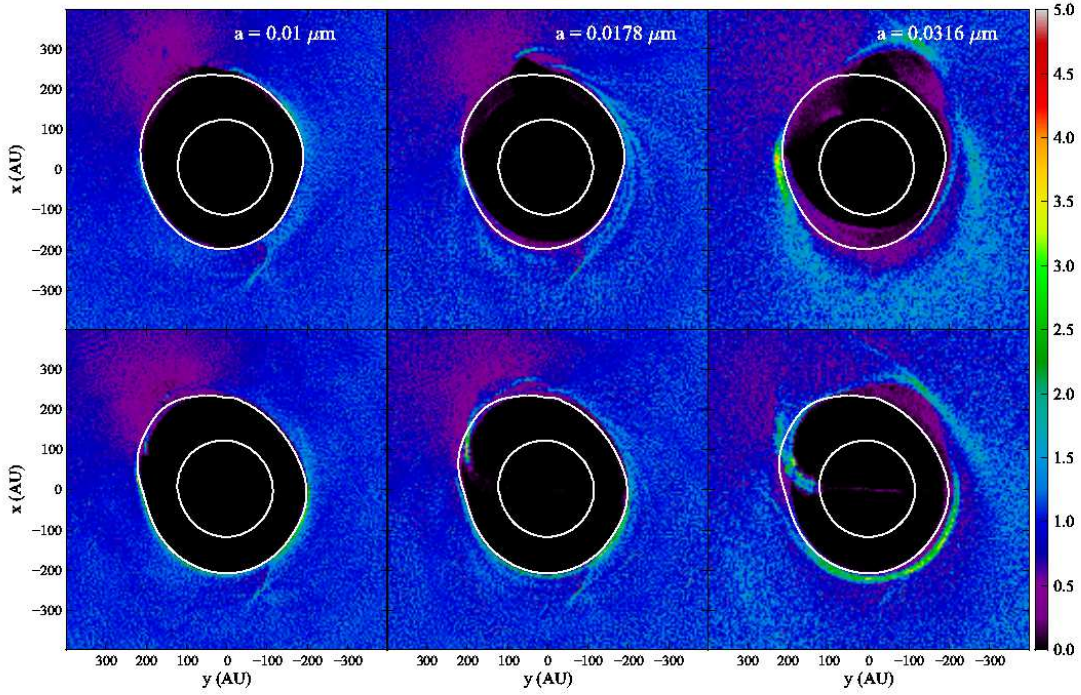}
\caption{Same as Figure \ref{fig:ddens_group1_x} except for slices
perpendicular to the ecliptic plane and roughly perpendicular to the
interstellar inflow direction.  The view orientation is looking downstream (in
the $-z$ direction). In this orientation the heliopause appears as the
somewhat lopsided white oval outside the more circular termination shock.}
\label{fig:ddens_group1_z}
\end{figure}

\begin{figure}[ht!]
\plotone{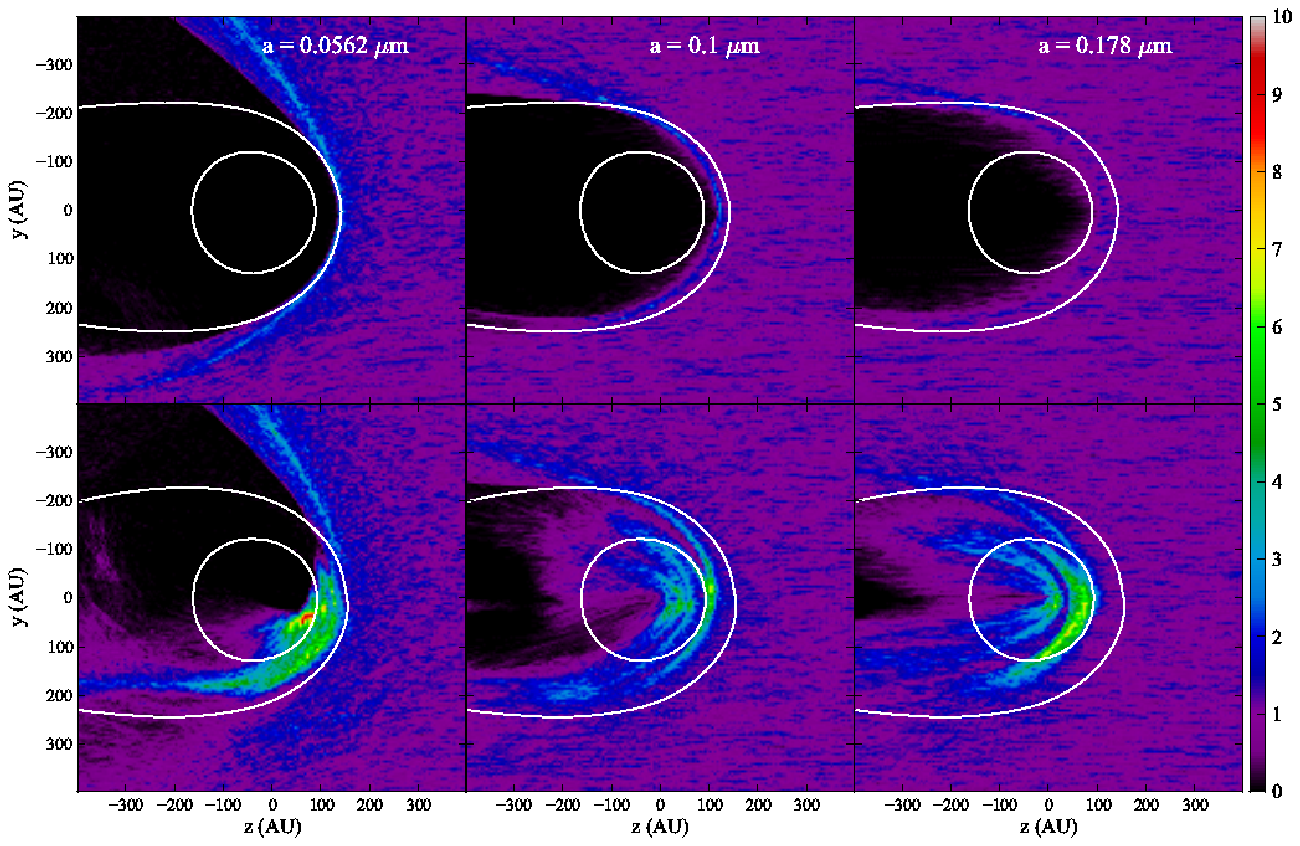}
\caption{Same as Figure \ref{fig:ddens_group1_x} except for 0.0562, 0.1 and
0.178 \micron\ grains (as labeled).}
\label{fig:ddens_group2_x}
\end{figure}

\begin{figure}[ht!]
\plotone{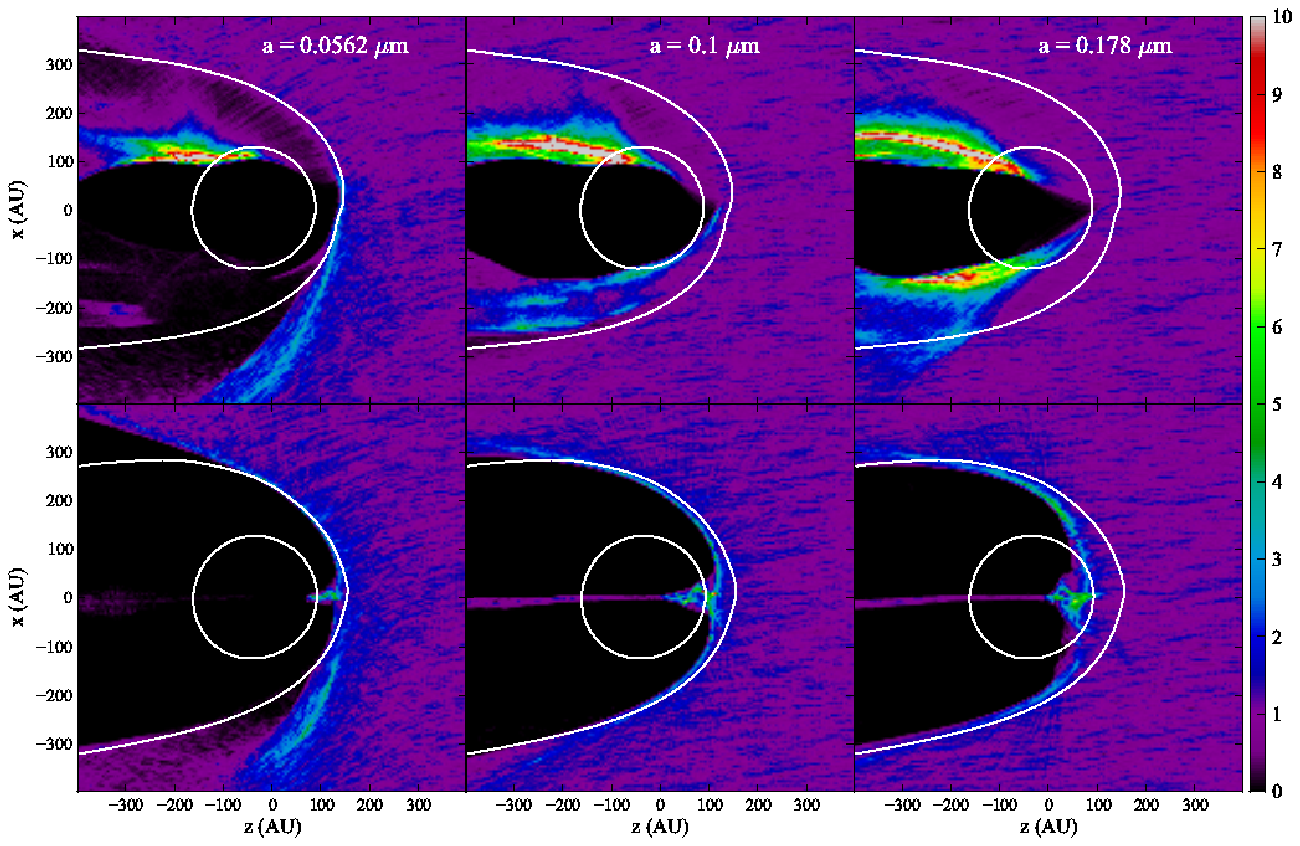}
\caption{Same as Figure \ref{fig:ddens_group1_y} except for 0.0562, 0.1 and
0.178 \micron\ grains (as labeled).}
\label{fig:ddens_group2_y}
\end{figure}

\begin{figure}[ht!]
\plotone{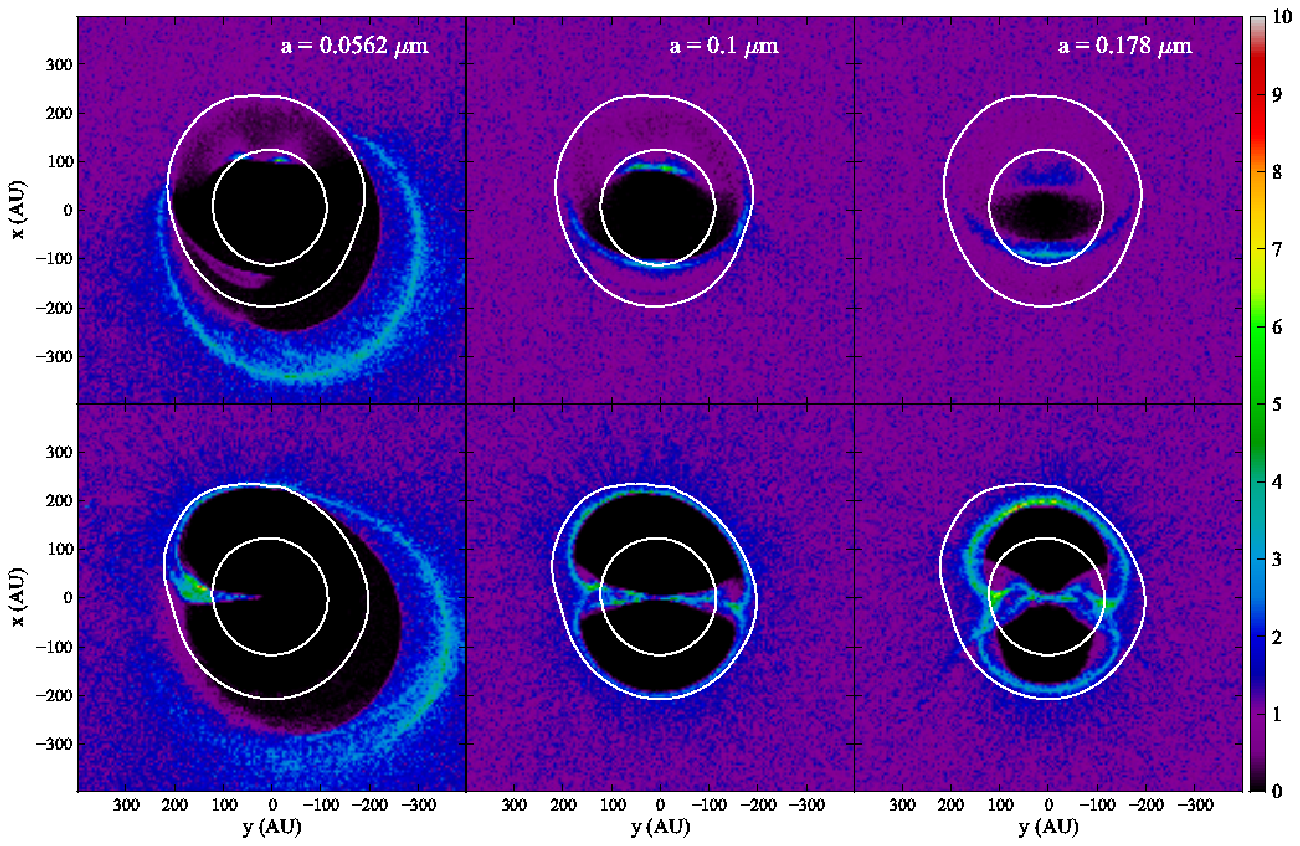}
\caption{Same as Figure \ref{fig:ddens_group1_z} except for 0.0562, 0.1 and
0.178 \micron\ grains (as labeled).}
\label{fig:ddens_group2_z}
\end{figure}

\begin{figure}[ht!]
\plotone{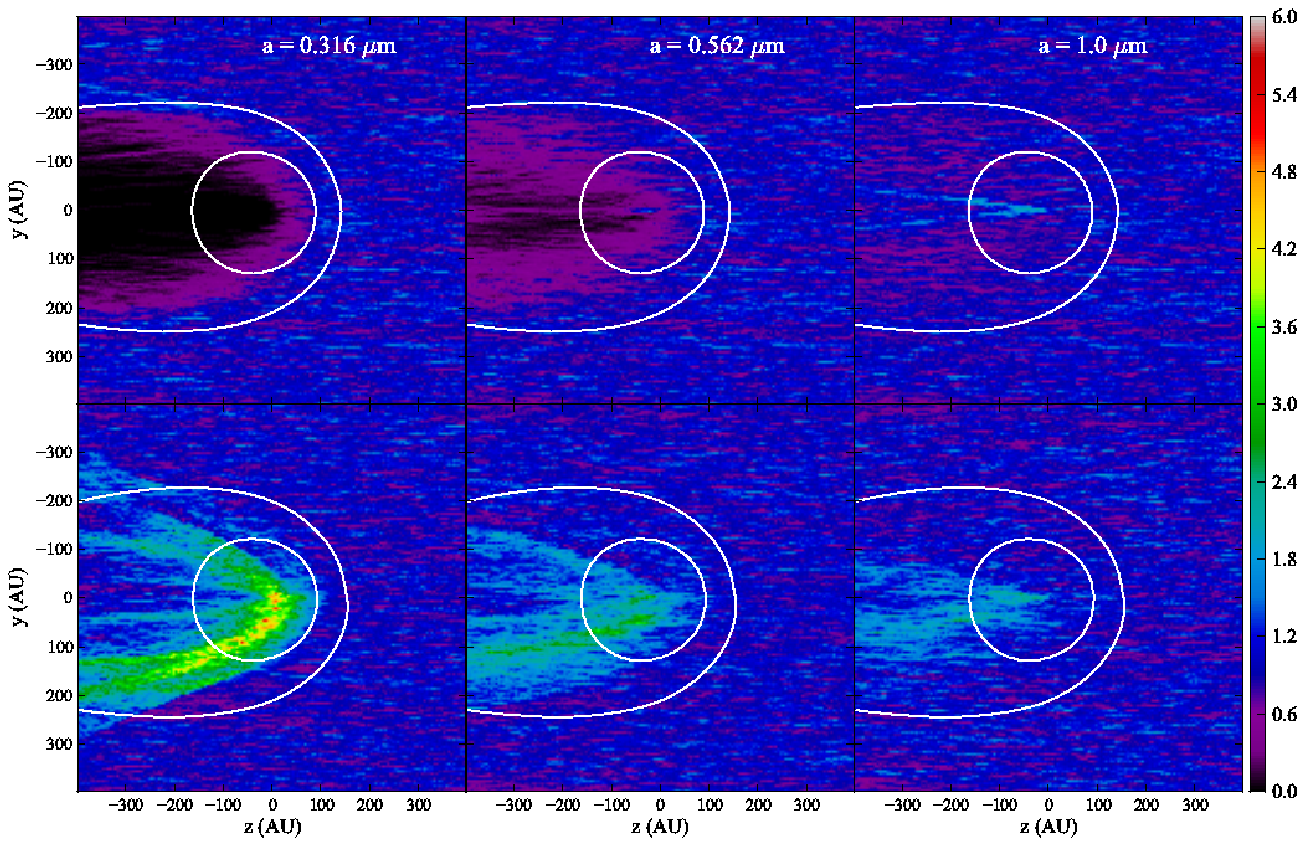}
\caption{Same as Figure \ref{fig:ddens_group1_x} except for 0.316, 0.562 and
1.0 \micron\ grains (as labeled).}
\label{fig:ddens_group3_x}
\end{figure}

\begin{figure}[ht!]
\plotone{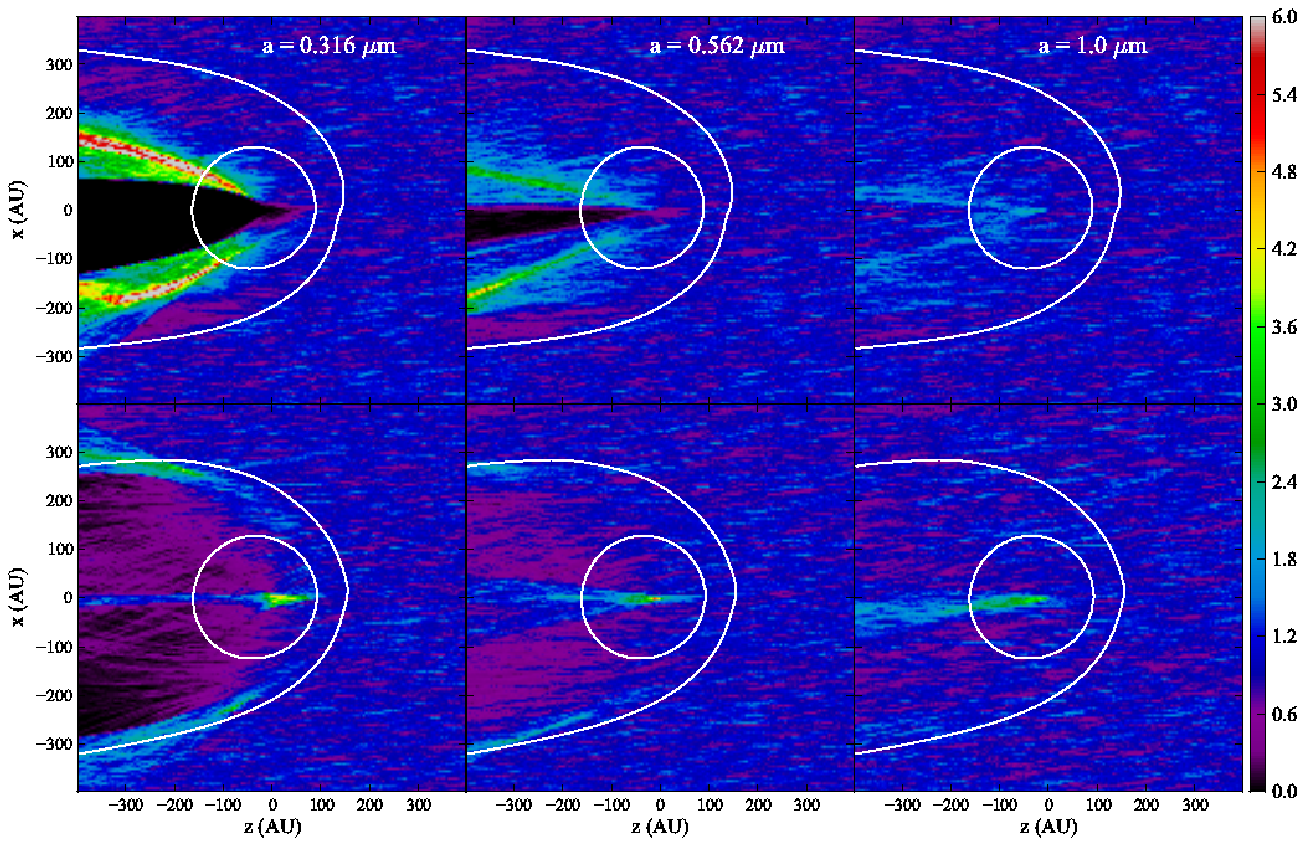}
\caption{Same as Figure \ref{fig:ddens_group1_y} except for 0.316, 0.562 and
1.0 \micron\ grains (as labeled).}
\label{fig:ddens_group3_y}
\end{figure}

\begin{figure}[ht!]
\plotone{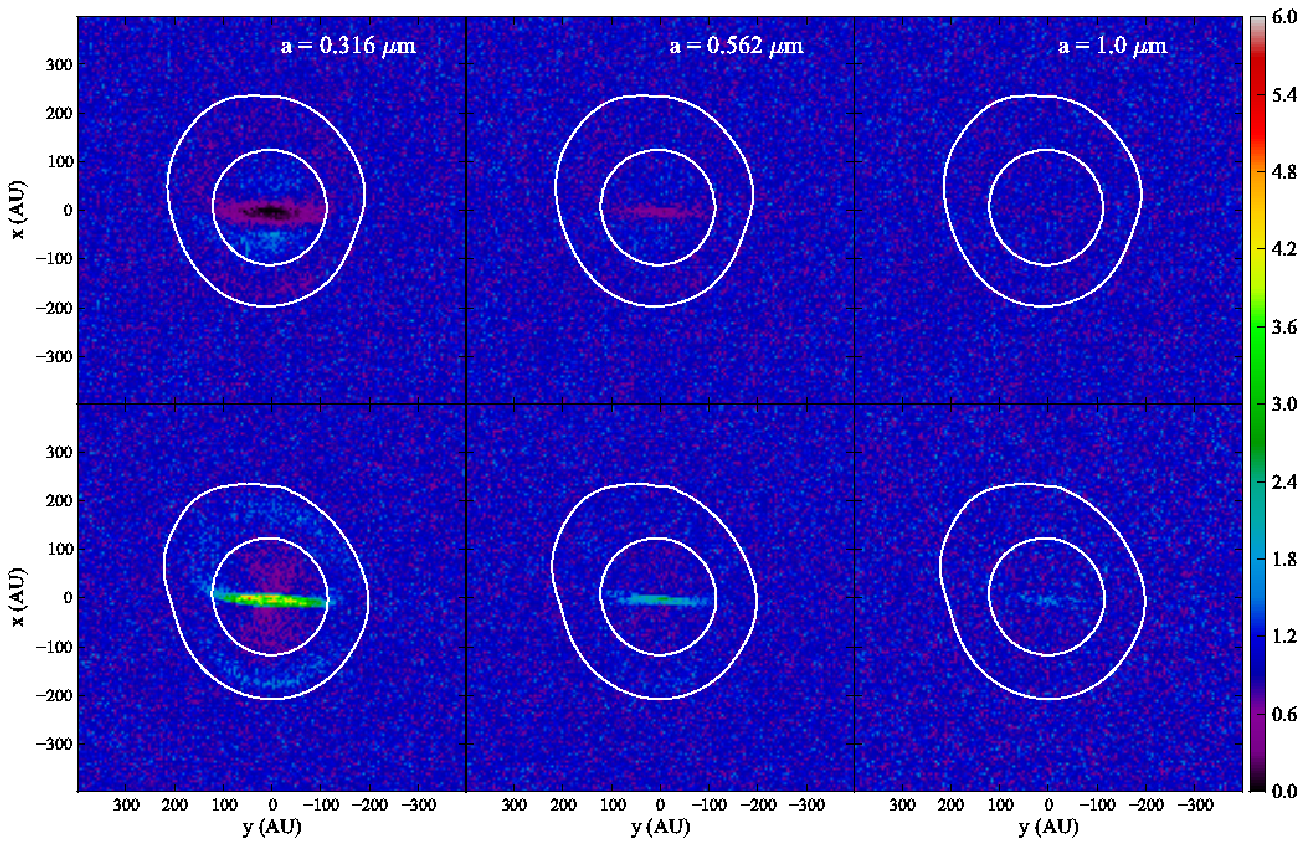}
\caption{Same as Figure \ref{fig:ddens_group1_z} except for 0.316, 0.562 and
1.0 \micron\ grains (as labeled).}
\label{fig:ddens_group3_z}
\end{figure}

The primary outputs of our calculations are the 3-D dust density
distributions.  In Figures \ref{fig:ddens_group1_x}, \ref{fig:ddens_group1_y}
and \ref{fig:ddens_group1_z} we show slices of the dust density relative to
the ambient value for the $z$-$y$, $z$-$x$, and $y$-$x$ planes respectively,
comparing results for the de-focusing solar wind magnetic field (SWMF)
orientation which has magnetic north (top row) with those for the focusing
SWMF polarity (bottom row).  These figures are for the smallest grain sizes
calculated, $a_\mathrm{grain} = 0.01$, 0.0178 and 0.0316 $\mu$m. In Figures
\ref{fig:ddens_group2_x}-\ref{fig:ddens_group2_z} we show the same type of
plots for larger grains, $a_\mathrm{grain} = 0.0562$, 0.1 and 0.178 $\mu$m and
in Figures \ref{fig:ddens_group3_x}-\ref{fig:ddens_group3_z} for the largest
grain sizes calculated, $a_\mathrm{grain} = 0.316$, 0.562 and 1.0 $\mu$m.  The
Sun-centered coordinate system used is described above
(\S\ref{sec:heliomodel}).  Here the plots are oriented such that the
heliosphere is viewed from above for the $z$-$y$ plots, where the  upwind
direction is to the right and ecliptic longitude increases in negative-y
directions. The $z$-$x$ plots show a meridian cut.  The  $y$-$x$ plots are
looking downstream from a viewpoint in the upstream interstellar medium.

An abundance of features are apparent in these model data, but a few stand out
immediately. First, it is clear that for the de-focusing SWMF polarity only
fairly large grains, $a_\mathrm{gr} \gtrsim 0.3$ \micron, can reach Earth's
orbit, though grains as small as 0.1 \micron\ can penetrate the termination
shock.  Thus it is hard to reconcile this model with the observed size
distribution, which includes grains as small as $\sim 0.05$ \micron.  For the
focusing polarity field, again grains as small as 0.1 \micron\ can penetrate
the termination shock, but in this case the focusing effects of the SWMF
allows such grains to reach all the way to the Sun, as they are concentrated
in the ecliptic plane. Indeed even grains as small as 0.02 \micron\ can reach
the Sun for the focusing polarity.  These strong differences in the
distributions of dust grains of a given size between the different magnetic
polarities persists, even up to 1 \micron\ grains, though for larger grains
the focusing/de-focusing effects of the SWMF are relatively small and
mitigated by the effects of gravitational focusing by the Sun.   Even for the
largest grains, $a = 1$ \micron, the dust focusing tail shows some solar cycle
dependence (e.g. Figures 10--12).  For both SWMF polarities, the dust
trajectories and thus dust density distributions are quite sensitive to grain
size.  This is because the gyroradii of the grains range from $< 1$ AU for
0.01 \micron\ grains to $\sim 10^5$ AU for 1 \micron\ grains at the
heliopause.

In the dust density figures it is clear that the smallest grains are
effectively completely excluded from the inner heliosheath, being diverted
along with the interstellar plasma around the heliopause.  At larger grain
sizes, more penetration of the heliopause and the effects of the larger
gyroradii can be seen. As an example, in Fig.\ \ref{fig:ddens_group1_y} in the
upper right panel ($a = 0.0316$ \micron), the low densities for large $x$
values above the heliopause show how grains that have been diverted do not
follow the heliopause, but gain a substantial $+x$ velocity that carries them
away from the heliosphere.  The regions with significant dust density inside
the heliopause in this panel are from scattering of grains in from the flanks
of the heliopause (because their gyroradii are large enough to allow them to
cross the heliopause).  The enhancement in grain density in the region just
upwind of the heliopause in the $z$-$y$ plot for this same grain size is also
caused by the decoupling.  The effects of decoupling of the grains from the
gas become much more pronounced for larger grain sizes.  For 0.0562 \micron\
grains the effects of scattering off the heliopause are clear in Figs.
\ref{fig:ddens_group2_x}-\ref{fig:ddens_group2_z}.  In the $z$ slices in
particular, it can be seen that the grains, which have a small initial $-x$
velocity (recall that the upwind direction is $\sim 5\degr$ above the $+z$
axis) are deflected by about 200 AU over the $\sim 150$ AU distance since
encountering the heliopause.  We note that this grain size is peculiar in our
results because for both the de-focusing and focusing SWMF models we find a
zero density for the voxels near the Sun, although for the focusing SWMF
smaller grains (0.0178, 0.0316 \micron) do produce non-zero density.  This
result, which we discuss more below, appears to be caused by the fact that
grains of this size have gyroradii at the location of encountering the
heliopause, that are roughly the same size as the heliopause.  They thus
decouple substantially from the gas, but are strongly diverted from their
inflow paths.  

\begin{figure}[ht!]
\plotone{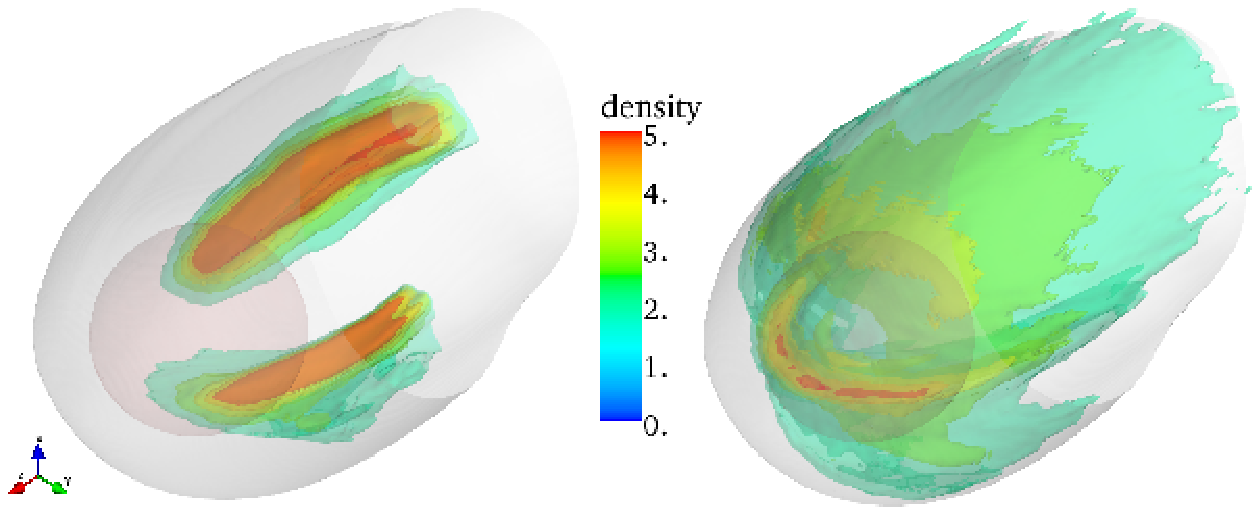}
\caption{Dust density enhancement for 0.178 \micron\ grains in the heliosphere
rendered as the 3-D surfaces corresponding to factors of 2, 2.75, 3.5, 4.25
and 5 over the ambient density.  The de-focusing SWMF case is shown on the
left and the focusing SWMF polarity case is shown on the right.  It can be
seen that the focusing SWMF leads to enhanced density in a crescent shaped
region near the ecliptic plane while the de-focusing field leads to
concentration of the dust at the north and south ecliptic poles. These
plumes of dust in the constant polarity models, should blend into each other
in the true heliosphere with its time-variable SWMF polarity. Note: the
interstellar inflow comes from the lower left and flows toward the upper right.
\label{fig:ddens_3d}}
\end{figure}

Smaller grains, while more tightly coupled to the plasma (via the magnetic
field), can scatter and leak into the heliosphere wherein they may be focused
towards the ecliptic for the focusing SWMF.  Larger grains, while even more
decoupled from the plasma, are less diverted from their paths and,
again for the focusing SWMF, have an enhanced dust density near the Sun.
As the grain size increase, grains in the range of $\sim 0.05 - 0.2$ \micron\
(Figs.\ \ref{fig:ddens_group2_x}--\ref{fig:ddens_group3_z}) show increasing
degrees of penetration of the heliopause and are focused either near the
ecliptic poles, for the de-focusing SWMF, or in the ecliptic, for the focusing
SWMF.  The 3-D grain density distribution for 0.178\micron\ grains is
illustrated in Figure \ref{fig:ddens_3d}, which shows a rendering of the 3-D
surfaces corresponding to dust density enhancements (relative to the
interstellar density) of 2, 2.75, 3.5, 4.25 and 5. Going to the largest grain
sizes (Figs.\ \ref{fig:ddens_group3_x}-\ref{fig:ddens_group3_z}) the grains
are less and less affected by the heliopause and at the largest size, 1.0
\micron, are primarily affected by the gravitational focusing by the Sun.

The individual grain trajectories can be quite complex, especially for the
intermediate grain sizes, $\sim 0.05 - 0.3$ \micron, for which the grains are
significantly diverted from their original path but have large enough
gyroradii to penetrate to the inner heliosphere.  
The scattering of the grains inside the heliosphere
is clear as even grains that start out on similar trajectories diverge
strongly from each other. This scattering is primarily caused by the fact that
the gyroradii are as large or larger than a typical length scale for variation
in the magnetic field direction so the field sampled by different grains on
their trajectories is substantially different and becomes more so as they
propagate closer to the heliopause and inner heliosphere.  The differences in
trajectories are also amplified by the increase in magnetic field strength
caused by compression in the heliosheath.  

\subsection{Grain Size Distribution \label{sec:mrn}}
One of the most important reasons for calculating grain trajectories through
the heliosphere starting from the undisturbed ISM is to allow us to infer the
initial interstellar grain size distribution given the size distribution
observed by the \emph{Ulysses} and \emph{Galileo} spacecrafts.  With accurate
3-D heliosphere models and grain trajectory calculations one could, in
principle, infer from the observations the interstellar size distribution,
though of course grain sizes that are either completely excluded from the
inner heliosphere or not observable because of instrumental limitations cannot
be constrained by this method.  In this study we have made progress toward
this goal by employing accurate, though non-evolving, heliosphere models.  As
we discuss further below, however, the lack of solar cycle evolution of the
heliosphere, in particular the SWMF, during the time the grains traverse the
inner heliosphere limits the applicability of our results. Assuming either a
focusing or de-focusing SWMF during an entire grain trajectory effectively
exaggerates the focusing or de-focusing the grains will experience.  We expect
that the truth will lie in between these extreme cases. The detailed behavior
of the grains with small to moderate gyroradii as the SWMF evolves is unclear
at present especially considering the complex magnetic topology caused
by the evolution and propagation of the heliospheric current sheets
\citep{Nerney_etal_1995}. Despite these limitations, we find below that
comparisons between our models and the in situ data provide valuable insights
into the properties of the grains and their interaction with the heliosphere.

Our calculations do not assume any particular size distribution (being carried
out independently for each grain size) since the grains are not assumed to
interact significantly with each other.  This assumption should be excellent
over the length scales relevant for the heliosphere, though in interstellar
shocks (with length scales hundreds of thousands of times longer), grain-grain
collisions are a major source of dust destruction \citep{Jones_etal_1994}.
The output of our calculations provide an unbiased result for each grain size,
since the dust density at each point in space is calculated relative to its
value in the ISM prior to interaction with the heliosphere.

Comparisons between the models and the \emph{in situ} observations are not
useful for the smallest grains or for regions of the model that are
infrequently populated with grains.  For the smallest grains, below the
detection threshold of the \emph{Ulysses} and \emph{Galileo} dust detectors
($m \approx 7\times10^{-16}$ g or $a \approx 0.04$ \micron\ for our assumed
grain density), we obviously have no information on their abundance in the
heliosphere.  Use of our theoretical data is also limited by our statistics:
if we have no counts in a voxel, this simply gives us an upper limit on our
predicted dust density relative to the initial interstellar dust density for
grains in that size range.  In the calculations that we present in this paper,
we expect, on average, $\sim 62$ counts per voxel (5 AU in each dimension) if
the dust density were undisturbed from its value in the ambient ISM.  Thus if
the dust density for a particular grain size is reduced by a factor of
$\gtrsim 62$ then we are likely to get no counts in the voxel and we are not
sensitive to how large the reduction in dust density is above that threshold.

Although our calculations use a regular grid of starting points for the
trajectories, they are, in some sense, similar to Monte Carlo
calculations.  Each individual trajectory is quite sensitive to its initial
conditions, so the starting positions, which have no \emph{a priori}
justification other than being sufficiently far upstream of the heliosphere,
could be considered to be random choices. A slight shifting of the starting
grid would lead to considerably different individual trajectories, though the
dust density results would be essentially unchanged. In addition, as mentioned
above, the initial gyrovelocity for each grain has a random direction other
than that is in the plane perpendicular to the magnetic field (i.e. 90\degr\
pitch angle).  This small addition of randomness to the initial velocities is
sufficient to remove most of the identifiable signature of the initial grid in
the final dust density results.  Thus we can treat the dust density results in
a statistical manner and assume the uncertainty in the ``true'' number of
counts we should get for each voxel is consistent with Poisson statistics. As
shown by \citet{Gehrels_1986}, for small numbers of counts the usual
$\sqrt{N}$ approximation is not a good approximation for the confidence limits
and we use instead his results to determine the statistical 1-$\sigma$ error
bars. The 1-$\sigma$ upper limits are $1 + N + \sqrt{N + \frac{3}{4}}$ and
lower limits are $N\left(1 - \frac{1}{9N} - \frac{1}{3\sqrt{N}}\right)^3$.
Thus for a voxel in with no counts, the upper limit is $1 +
\sqrt{\frac{3}{4}}$.

\begin{figure}
\plotone{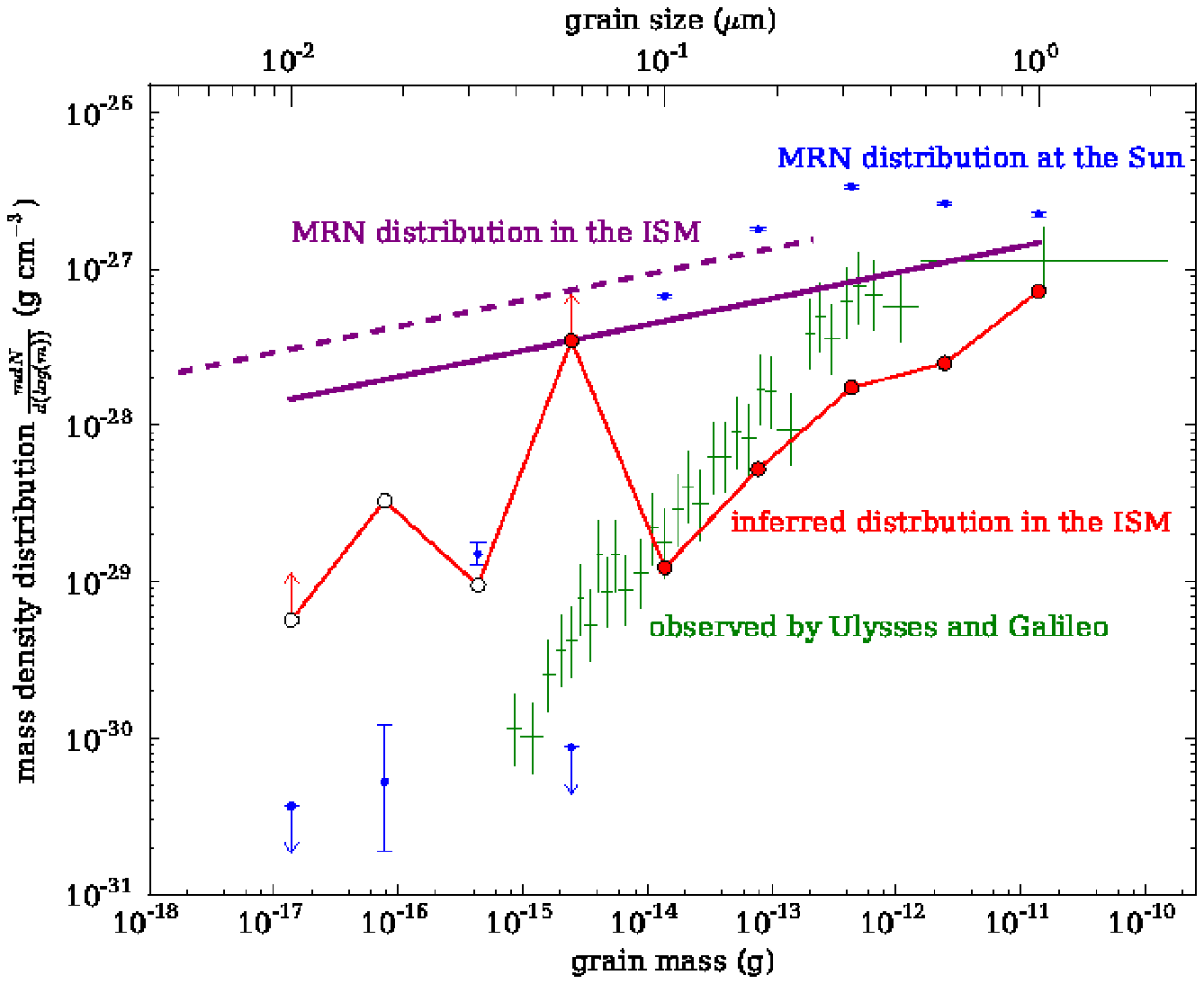}
\caption{Size distribution of interstellar grains in the ISM and the
heliosphere (plotted as mass density distributions).  In green are the
\emph{Ulysses} and \emph{Galileo} data where the bin widths are set so as to
get the same total number of counts per bin \citep[see][]{Frisch_etal_2009}.
Plotted in purple are the standard \citet{Mathis_etal_1977} size distribution
which cuts off at $a_\mathrm{gr} = 0.25$ \micron\ (dashed) and the same power
law but extended to 1 \micron.  Those size distributions are both normalized
so as to produce a
gas-to-dust mass ratio of 150.  In blue is plotted the size distribution that
would be observed near the Sun if the interstellar size distribution were the
extended MRN and the focusing SWMF polarity model applied.  The error bars
here are statistical.  The red line shows the inferred interstellar grain size
distribution given the observed size distribution and the modeled grain
transport (again for the focusing SWMF polarity).  The empty circles at the
small grain size end are based on an extrapolation of the observed grain size
distribution below the detection threshold.
\label{fig:size_distn}}
\end{figure}

To scale the results of our models to actual densities in the ISM or
heliosphere we need to assume an initial grain size distribution including the
scaling relative to the gas density.  Differences between the initial grain
distribution at ``infinity'' and the observed \emph{in situ} dust distribution
should then be explained in principle  by grain propagation models.  In Figure
\ref{fig:size_distn} we show the data from \citet[and references
therein]{Frisch_etal_2009} for \emph{Ulysses} and \emph{Galileo} dust
detection along with models and inferred grain size distributions.  Here,
rather than the grain size distribution, $\frac{dn}{da}$, we plot the grain
mass per logarithmic mass bin as was done in \citet{Frisch_etal_1999} and
\citet{Frisch_etal_2009}.  Plotted this way, it is apparent that most of the
mass is in grains on the large end of the scale.  In the figure, we plot the
standard MRN power-law distribution for grain sizes \citep[dashed, purple
curve]{Mathis_etal_1977} which extends from 0.005 \micron\ to 0.25 \micron\
with the normalization set by the assumption of an interstellar density of
$n_\mathrm{H} = 0.22$ cm$^{-3}$ and a gas-to-dust mass ratio of 150.  The H
density and gas-to-dust ratio are taken from the LIC photoionization models of
\citet{Slavin+Frisch_2008}, which are successful in that they predict
interstellar boundary conditions for the heliosphere that are consistent with
local ISM parameters derived from heliosphere models.  Since the upper size
cutoff for the standard MRN size distribution is well below the size of the
largest grains observed (for silicate grains; graphite grains sizes up to 1
\micron\ are allowed), we also plot an MRN-type size distribution but with
higher lower and upper size limits ($0.01 - 1.0$ \micron), and using the same
$n_\mathrm{H}$ and dust-to-gas ratio for normalization, that we refer to as a
``shifted MRN'' size distribution.  We note that more recent detailed ISM
grain models such as \citet{Zubko_etal_2004} have grain size distributions
that differ substantially from a simple power-law and in some cases extend
significantly above the 0.25 \micron\ cutoff of the MRN model.  However, no
proposed ISM grain model has a size distribution similar to that observed in
the heliosphere \citep{Draine_2009}, because such a size distribution is
inconsistent with observed extinction curves.  We note, however, that the 
extinction curves that such grain models aim to match are derived from
observations over kiloparsec long lines of sight. This implies that either the
size distribution in the LIC is very atypical of the ISM or it has been
substantially distorted by the transit of the grains through the heliosphere.
Our results provide insights into the effects of those distortions.

If we assume that the shifted MRN distribution, represents the grain size
distribution in the undisturbed ISM, then, using the focusing SWMF polarity
heliosphere model, we find that we would expect the detections shown as the
blue points in Figure \ref{fig:size_distn}. For these points we use the
density results for the voxels consistent with the orbit of \emph{Ulysses},
including ranges in $x$, $y$ and $z$ of $-3$ to $+3$ AU, $-1.3$ to $+5.4$ AU
and $-2.1$ to $+2.1$ AU respectively (using the coordinate system described in
\S\ref{sec:heliomodel} above).  The points are generally well above the actual
detections except for the point at 0.0562 \micron, for which no counts are
found in the model for the region close to the Sun (see discussion below).
This suggests that the focusing model provides too much focusing (at least for
the assumed size distribution), as would be expected if the grains experienced
a de-focusing phase as well as focusing during their transit of the
heliosphere.  If instead we use the actual observations (interpolated to get
the grain densities at the calculated grain sizes) then we can infer the size
distribution in the interstellar medium by using the model
enhancement/depletion of grains caused interaction with the heliosphere.  We
have done that, using spline interpolation on the data, with the results
plotted as the red points in the figure.  These points are correspondingly
well below the shifted MRN model line (again with the exception of the 0.0562
\micron\ grains).  In going from the observed points to the size distribution
in the undisturbed ISM we divide by the enhancement factor, i.e. ratio of dust
density for grains of a given size to that in the undisturbed ISM.  Therefore
a larger enhancement leads to a lower value inferred for the ISM.  Thus the
model with the focusing SWMF polarity appears to over-predict the focusing and
predicts a very small dust-to-gas ratio in the LIC.

At the low end of the size distribution we have done a linear (in the log,
i.e.\ powerlaw) extrapolation of the data down from the $\sim 0.1$ \micron\
detection threshold of \emph{Ulysses} to 0.01 \micron\ to explore the
implications for the grain size distribution.  If the grain size distribution
near the Sun really does extend in this way to smaller sizes, then the model
results imply a flattening of the size distribution in the ISM for for these
small grain sizes.  Since we there is no data on interstellar grains of those
sizes in the heliosphere, this result (which also only applies for the
focusing-only SWMF) is purely speculative at this point.

If we instead interpret the dust measurements using the de-focusing SWMF
polarity model the results are starkly different.  Since all but the three
largest grain sizes modeled (0.316, 0.562 and 1.0 \micron) produce no counts
for the voxels near the Sun, the inferred distribution of grains in the ISM
has, except for grain sizes $\geq 0.316$ \micron, very high lower limits that
generally exceed the extended MRN distribution.  Such a high density of dust
would demand a small gas-to-dust mass ratio, $R_{g/d} < 38$, and would
conflict with limits on total cosmic abundances for the elements that make up
the dust.  On the other hand the results for the focusing SWMF polarity model
(discussed above) implies a rather small dust density and thus a high
gas-to-dust ratio, $R_{g/d} \sim 820$ (ignoring the 0.0562 \micron\ grain
size).  This would indicate a high degree of dust destruction in the ISM and
only a small mass of heavy elements tied up in the dust.  This again conflicts
with observations, in particular those that indicate significant gas phase
depletion of several elements, e.g.\ Fe, Mg and Si, from the gas phase
\citep{Slavin+Frisch_2008}.  

The fact that neither the focusing SWMF nor de-focusing SWMF models can be
comfortably accommodated with our information on gas phase elemental
abundances in the LIC points to the limitations of constant polarity models
for the SWMF.  This is not surprising since the grains require $\sim 20$ years
to travel between the termination shock and inner heliosphere.  The fact that
the predictions of our two heliosphere models lead to bracketing of the likely
possibilities for the inferred gas-to-dust ratio and grain size distribution
suggests that grain trajectory modeling needs to include the time variation of
the solar wind, and its magnetic field polarity in particular, over the solar
cycle.  The strong dependence of the grain density on the SWMF polarity shows
that the recovery of the mass distribution of interstellar dust from \emph{in
situ} measurements requires a complete understanding of the effect of the
solar cycle on the global heliosphere, and grain trajectory models that
incorporate interactions with the 3-D global heliosphere.

We note that the grains with the particular size of 0.0562 \micron\ are
predicted to have very low density (zero counts calculated) near the Sun for
both heliosphere models, though for the focusing SWMF model a substantial
amount of dust penetrates the heliopause.  As discussed in \S \ref{sec:results}
above, this appears to be caused by the size of the gyroradius for grains of
this size, which is large enough that the grains decouple from the gas enough
to penetrate the heliopause and termination shock, but small enough that their
path is substantially diverted from its initial direction.  We expect that
additional perturbations to the initial conditions for the grains and
especially spatial and temporal variations in the solar wind will wash out
this feature leading to a relatively featureless grain size distribution such
as is observed.

The heliosphere models that we use in this paper are consistent with all the
information available up until recently on the ISM and the heliosphere.  This
includes the speed of the inflowing ISM and the orientation of the
interstellar magnetic field required to produce the asymmetry in the shape of
the termination shock as revealed by the location of the \emph{Voyager} 1 and
2 crossings.  Recent data from IBEX  \citep{Moebius_etal_2012} indicate that
the speed of the interstellar inflow is about 12\% less than previously
determined, $\sim 23$ km s$^{-1}$ rather than 26.4 km s$^{-1}$.  This lower
velocity appears to to rule out a bow shock ahead of the heliosphere
\citep{McComas_etal_2012}.  If this result holds up, then other parameters of
the heliosphere models, e.g.\ interstellar magnetic field strength, will
likely require some small adjustments as well.  We do not expect this to make
any difference regarding the fundamental conclusions of this paper, especially
since the models that we have used already do not have a shock transition
between the ISM and the outer heliosheath.

\section{Conclusions \label{sec:conclusion}}

We have presented the first simulations of interstellar grain propagation
through the heliosphere that have incorporated a realistic 3-D global
heliosphere model that accommodates recently gained knowledge on the shape and
size of the heliosphere, including the asymmetries due to the large angle
between the interstellar gas flow and ISMF direction.  While our understanding
of the heliosphere continues to improve, the heliosphere models used in
this study capture the essential characteristics of its shape and the
characteristics that affect distribution of ISD in the heliosphere (except
for its time variability during grain propagation).  

Our models include detailed calculations of the grain charging based on a
standard interstellar grain model, using olivine silicates with density 3.3 gr
cm$^{-3}$, and including a realistic solar radiation field.  The calculations
follow the grain trajectories that originate in the undisturbed interstellar
medium well outside of the heliosphere.  Our results indicate that inferences
on the grain size distribution and abundance in the interstellar medium
surrounding the heliosphere depend sensitively on the heliosphere model and in
particular on the polarity of the solar wind magnetic field.  For a focusing
polarity of the field, grains over a wide size range can penetrate the
heliopause and are focused in the ecliptic.  For de-focusing polarity, only
the larger grains we studied, $\gtrsim 0.3$ \micron, are found to create
non-zero density at the Sun.  For either SWMF polarity, the largest grains
($\sim 1$ \micron) have enhanced dust density relative to that in the ISM,
because of gravitational focusing, which helps to explain the larger than
expected observed density of large grains in the inner heliosphere near the
Sun.

Grains in the size range near 0.2 \micron\ are diverted into dust plumes along
the flanks of the heliosphere inside of the termination shock, with the plumes
located in the polar regions for the de-focusing polarity model and in the
ecliptic region for the focusing polarity model.  In the true time-variable
heliosphere, these dust grains will sample both polarities of the SWMF and the
different magnetic morphologies and solar wind conditions that will exist over
the course of the solar cycle.  We speculate that this may lead to these dust
density enhancements being smeared into some shape in between these extremes,
perhaps an asymmetric dust shell inside of the termination shock but somewhat
upstream of the Sun.  Such a feature has the potential to create an
unaccounted for contribution to the infrared and microwave sky background.
This possibility requires further investigation.

The total dust density inferred for the ISM, using our models, is either too
small, for the focusing polarity, or too large, for the de-focusing polarity
to be consistent with the gas phase abundances inferred from absorption line
data \citep[e.g.,][]{Slavin+Frisch_2008}.  This points to the need to
use heliosphere models that include the time dependence of the solar
wind magnetic field over the course of a solar cycle in grain trajectory
calculations.  Our calculations, by using a single polarity for the SWMF over
the decades long course of a grain trajectory, effectively bracket the
possible outcomes for the grain density inside the heliosphere.  With future
calculations that include the time evolution of the SWMF we hope to narrow the
range of predicted grain densities, leading to a robust inference on the
interstellar grain size distribution from \emph{in situ} observations of
interstellar dust in the heliosphere.

\acknowledgements
This work has been supported by NASA grant number NNX08AJ33G to the University
of Chicago, and by the Interstellar Boundary Explorer mission as a part of
NASA's Explorer Program. HRM acknowledges support by NASA SHP SR\&T grant no.\
NNX10AC44G.  The UAH team was supported by NASA grants NNX09AG63G and
NNX12AB30G, and DOE grant DE-SC0008334.

\appendix

\section{Individual Grain Trajectories}
\begin{figure}[ht!]
\plotone{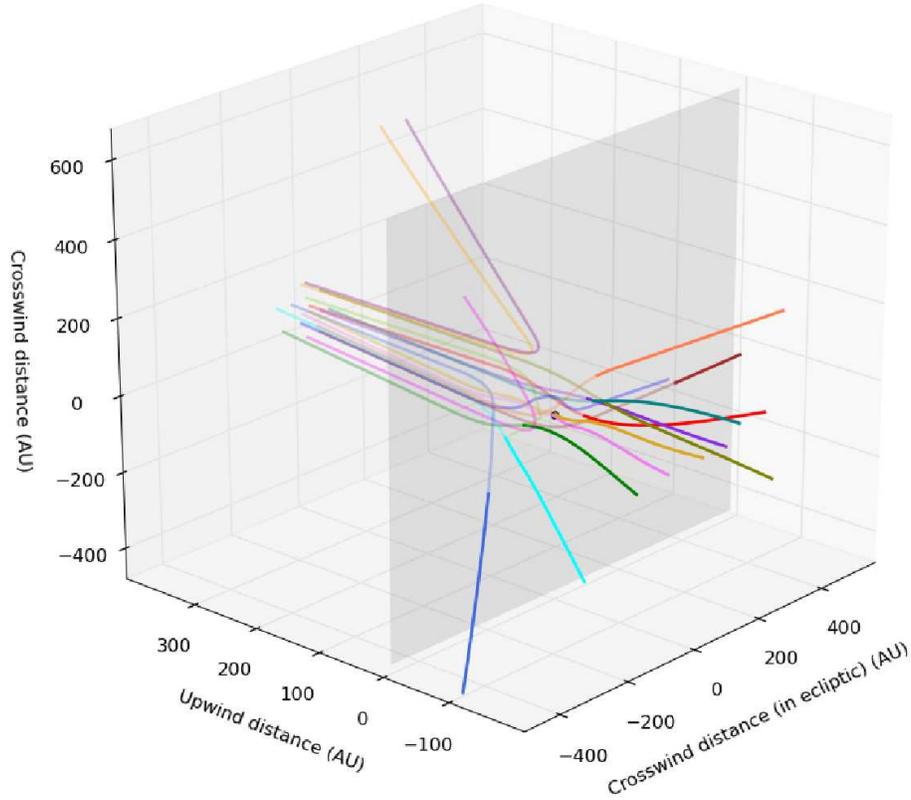}
\caption{Trajectories for 0.178 \micron\ grains for focusing polarity.
Included in the plot are just the parts of the trajectories with $z$ between
$-200$ and $+400$ AU.  The color coding was done as a way to help trace the
trajectories.  Also the trajectories are transparent when they are upwind and
once they cross the $z = 0$ plane, indicated by the transparent plane, they
are made opaque. The Sun's location is indicated by a circle.  Note the strong
level of scattering experienced by some grains, while for others their
trajectories are relatively straight.  These trajectories are a small subset
of the more than $10^6$ trajectories calculated to make the dust density
plots.}
\label{trajs}
\end{figure}

Examination of trajectory groups in the model calculations shows the
complexity of individual trajectories.  In Figure \ref{trajs} we
show a group of 16 trajectories for 0.178 \micron\ grains that result in
relatively close passage to the Sun to illustrate more clearly the way in
which grains are scattered in the inner heliosphere.  Gravity and radiation
pressure do not have a strong influence on these trajectories though for
somewhat larger grains, gravity does play an important role for grains passing
near the Sun.  Some trajectories are dominated by a single
scattering due to the Lorentz force, while others experience several
scatterings.  The small initial gyrovelocity given to the grains (due to
interstellar turbulence) acts to create a stochastic element to the
trajectories which is enhanced by their subsequent interactions with
heliosphere.

\end{document}